\documentclass[pra,english,showpacs,preprintnumbers,amsmath,amssymb,nofootinbib,
twocolumn,superscriptaddress]{revtex4-1}
\usepackage{graphicx}
\usepackage{epstopdf}
\usepackage{epsfig}
\usepackage{bm}
\pdfoutput=1

\usepackage[latin1]{inputenc}
\usepackage{color}
\usepackage{amssymb}
\usepackage{amsmath}
\usepackage{tabularx}
\def\0#1#2{\frac{#1}{#2}}

\def\s0#1#2{\mbox{\small{$ \frac{#1}{#2} $}}}

\def\CZ{{\mathcal Z}}
\def\CD{{\mathcal D}}

\newcommand{\beq}{\begin{equation}}
\newcommand{\eeq}{\end{equation}}
\newcommand{\bea}{\begin{eqnarray}}
\newcommand{\eea}{\end{eqnarray}}
\newcommand{\tr}{\mathrm{tr}}

\newcommand{\pderiv}[2]{\frac{\partial #1}{\partial #2}}

\usepackage{babel}
\makeatother
\begin{document}

\title{Few-fermion systems in one dimension: Ground- and excited-state energies and contacts}

\author{L. Rammelm\"uller}
\affiliation{Fakult\"at f\"ur Physik, Technische Universit\"at Wien, 1040
Vienna, Austria}
\affiliation{Department of Physics and Astronomy, University of North Carolina,
Chapel Hill, North Carolina 27599-3255, USA}

\author{W. J. Porter}
\affiliation{Department of Physics and Astronomy, University of North Carolina,
Chapel Hill, North Carolina 27599-3255, USA}

\author{A. C. Loheac}
\affiliation{Department of Physics and Astronomy, University of North Carolina,
Chapel Hill, North Carolina 27599-3255, USA}

\author{J. E. Drut} 
\affiliation{Department of Physics and Astronomy, University of North Carolina,
Chapel Hill, North Carolina 27599-3255, USA}

\date {\today}

\begin{abstract}
Using the lattice Monte Carlo method, we compute the energy and
Tan's contact in the ground state as well as the first excited state of few- 
to many-fermion systems in a one-dimensional periodic box.
We focus on unpolarized systems of $N = 4,6,...,12$ particles, with a zero-range interaction, and 
a wide range of attractive couplings.  In addition, we provide extrapolations to the 
infinite-volume and thermodynamic limits.
\end{abstract}

\pacs{03.50.Fk, 67.85.Lm, 74.20.Fg}

\maketitle

%%%%%%%%%%%%%%%%%%%%%%%%%%%%%%%%%%%%%%%%%%%%%%%%%%%%%%%%%
\section{Introduction}
About a decade ago, in a remarkable series of papers, Tan showed that
high-momentum correlations, in few- and many-body quantum systems with short-range 
interactions, are governed by a single number: the contact $C$~\cite{TanContact}.
In particular, he found analytically that the high-momentum tail of the momentum
distribution obeyed the law $n(k)\sim C/k^4$, and that $C$ also governed the behavior of the energy upon small adiabatic
changes in the coupling strength (as driven, e.g., by an external magnetic field in ultracold atom experiments~\cite{RevExp,
RevTheory}). These developments took place more or less in parallel with the work of
others (see, e.g., Ref.~\cite{Leggett}), and they were eventually understood in terms of the operator-product expansion of quantum
field theory by Braaten and colleagues~\cite{BraatenEtAl}. Other works followed that considered the high-momentum behavior of more complex
quantities (such as the stress energy tensor), which led to the appearance of the contact in sum rules involving the shear and
bulk viscosities, the superfluid density, and other response functions~\cite{RanderiaTaylor, SonThompson, Hofmann}.

While those advances were by and large analytic, the calculation of $C$
itself in few- or many-body systems typically requires solving those problems numerically, e.g., via quantum Monte Carlo
methods. The large-momentum and adiabatic relations mentioned above (see also Refs.~\cite{Werner,Valiente}), however, provide possible 
avenues for the computation of $C$, which will in general depend on all the dynamic and thermodynamic variables of the system.
Calculations of $C$ using nonperturbative methods (Bethe ansatz in one dimension, lattice and diffusion Monte Carlo methods in two and three dimensions) 
appeared in Ref.~\cite{DLT} in three dimensions, in Ref.~\cite{BertainaZhang} in two dimensions, and in Ref.~\cite{BarthZwerger} in one dimension. Excellent 
reviews can be found in Refs.~\cite{WernerCastin, BraatenReview, ContactReview}.

In this work, we present our calculations of the ground- and first-excited-state energy and Tan's
contact for multiparticle nonrelativistic fermion systems in a one-dimensional (1D) box with periodic boundary conditions (i.e., a ring).
We cover the range from few- to many-body systems and couplings from noninteracting to strongly coupled.
On the experimental side, this problem has been explored in Ref.~\cite{Zurn}, although that realization differs from
the present system as it features an external harmonic trapping potential.
On the theory side, the problem can be solved using the Bethe ansatz technique~\cite{BetheAnsatzReview}, but we have chosen to approach it
using lattice Monte Carlo methods (i.e., lattice field theory formulations powered computationally by lattice-QCD techniques). 
The latter are applicable in situations where the former is not, notably in higher dimensions and in the presence of
an external potential. Furthermore, lattice methods are an area in which remarkable strides are 
currently being made across physics: from materials science~\cite{GrapheneLattice} to nuclear physics~\cite{LeeEtAl},
and of course lattice QCD~\cite{LQCDReviewAndBooks}. Therefore, understanding the advantages, disadvantages, and potential
of these methods is both interesting and timely.

%%%%%%%%%%%%%%%%%%%%%%%%%%%%%%%%%%%%%%%%%%%%%%%%%%%%%%%%
\section{Hamiltonian, scales, and dimensionless parameters} 
As anticipated above, we focus on a 1D system of nonrelativistic fermions with a zero-range 
interaction (also known as the Gaudin-Yang model~\cite{GaudinYang}) and periodic boundary conditions,
such that the Hamiltonian is given by
\beq
\hat H = -\frac{\hbar^2}{2m}\sum_i \nabla_{i}^2 - \sum_{i < j}
g\delta(x^{}_i-x^{}_j),
\eeq
where the sums are over all particles. We restrict ourselves to a two-species unpolarized system, 
but higher degeneracies can be studied with the same methods.
We employed the same technique as in Refs.~\cite{ImprovedActionsDrut} adapted to one dimensions.
Following that approach, we placed our system in a Euclidean spacetime lattice of
extent $N^{}_x \times N^{}_\tau$ and used a 
Trotter-Suzuki decomposition of the Boltzmann weight followed by a
Hubbard-Stratonovich transformation~\cite{HS}. 
The path integral is evaluated using Metropolis-based Monte Carlo methods (see,
e.g., Ref.~\cite{MCReviews}), specifically in the form of the hybrid Monte Carlo algorithm.
Further details on the method employed here are explained below.

In the following, we use units such that $\hbar = m = k_B = 1$, where $m$ is the
mass of the fermions.
The physical input parameters are the total particle number
$N=N_\uparrow^{}+N_\downarrow^{}$, the size of the box $L=N^{}_x \ell$ 
(where $\ell = 1$ to set the length and momentum scales), and the (attractive)
coupling strength $g$;
only the last two of which are dimensionful. We use all of these to form one
dimensionless intensive quantity: the dimensionless coupling $\gamma$ given by 
\beq
\gamma = g / n, 
\eeq
where $n = N/L$ is the particle-number density, as is common in other 1D ground-state 
studies (see, e.g., Refs.~\cite{Tokatly, FuchsRecatiZwerger}).
The extent of the temporal direction is $\beta = \tau N^{}_\tau$, which we vary in
order to extrapolate to the large-$\beta \varepsilon^{}_F$ 
limit, where $\varepsilon_F^{} = k_F^2/(2m)$ and $k_F^{} = n \pi/2$; we further elaborate on the relevant 
scales below. Note that, in one dimension, fermions with a contact interaction are 
ultraviolet-finite, and therefore the bare coupling has a well-defined physical
meaning, namely, $g = 2/a^{}_0$, where 
$a^{}_0$ is the scattering length for the symmetric channel (see, e.g.,
Ref.~\cite{ScatteringIn1D}).

%%%%%%%%%%%%%%%%%%%%%%%%%%%%%%%%%%%%%%%%%%%%%%%
\section{Many-body method}

The (unnormalized) ground state of a many-body quantum system with Hamiltonian $\hat H$ can be obtained as
the large-$\beta$ limit of
\beq
| \psi \rangle_\beta^{} = \exp\left(-\beta \hat{H}\right) \,| \psi^{}_0 \rangle,
\eeq
as long as the ``guess'' state $| \psi^{}_0 \rangle$ has a nonvanishing projection onto the
true ground state $| \psi \rangle$.
Thus, one may write the ground-state expectation value of an operator $\hat O$ as
\beq
\langle \hat O \rangle = \lim_{\beta\to\infty} O_\beta,
\eeq
where we have defined
\beq
O_\beta \equiv \frac{\langle \psi^{}_0 |
\,\hat{U}(\beta,\beta/2)\,\hat{O}\,\hat{U}(\beta/2,0)\,| \psi^{}_0 \rangle}
{\langle \psi^{}_0 | \, \hat{U}(\beta,0) \,| \psi^{}_0 \rangle}
\eeq
and the imaginary-time evolution operator
\beq
\hat{U}(\tau_b,\tau_a) \equiv \exp\left[-(\tau_b-\tau_a)\hat{H}\right].
\eeq

In this work we will use this simple formalism and take
$|\psi^{}_0\rangle$ to be a Slater determinant of single-particle orbitals $\{\phi^{}_k\}$ given by
plane waves, where $k = 1,2,\dots, N^{}_\uparrow$ with $N^{}_\uparrow = N^{}_\downarrow = N/2$ 
being the number of fermions of each flavor.

We approximate the operator $\hat{U}$ using a Suzuki-Trotter decomposition
\beq
\hat{U}(\tau_a+\tau,\tau_a) =e^{-\tau\hat{T}/2}e^{-\tau\hat{V}}e^{-\tau\hat{T}/2}+O(\tau^3),
\eeq
where $\tau$ is our imaginary-time discretization parameters, and we have split the Hamiltonian 
into the one-body kinetic energy operator $\hat{T}$ and the two-body, zero-range interaction $\hat{V}$.  
At each time step $t$, we implement an auxiliary field transformation writing (generically)
\beq
e^{-\tau\hat{V}} = 
\int \CD \sigma(x)
e^{-\tau \hat{V}^{}_{\uparrow,\sigma}}e^{-\tau \hat{V}^{}_{\downarrow,\sigma}}
\eeq
where the $\hat{V}^{}_{s,\sigma}$ are one-body operators that depend on the
Hubbard-Stratonovich field $\sigma(x)$ and $\int \CD \sigma(x)$ is a sum
over all possible configurations of $\sigma$ at the specific time slice $t$.
After collecting the factors corresponding to each time step, we identify the
zero-temperature partition sum
\beq
\label{Eq:Z}
\CZ \equiv {\langle \psi^{}_0 | \, \hat{U}(\beta,0) \,| \psi^{}_0 \rangle} = \int \CD \sigma (x,t)\, P[\sigma],
\eeq
where now the path integral is over a spacetime varying field $\sigma (x,t)$, and we defined the Monte Carlo probability
\beq
\label{Eq:PsigmaDef}
P[\sigma] \equiv \langle \psi^{}_0|\,\hat{U}_\sigma(\beta,0)\, |\psi^{}_0 \rangle = {\det}^2 \left[{M}_\sigma(\beta) \right],
\eeq
where 
\beq
\hat{U}_\sigma(\tau^{}_a+\tau,\tau^{}_a) \equiv
e^{-\tau\hat{T}/2}e^{-\tau\hat{V}_{\uparrow,\sigma}}e^{-\tau\hat{V}_{\downarrow, \sigma}}e^{-\tau\hat{T}/2},
\eeq
and the matrix ${M}_\sigma(\beta)$ is the single-particle representation of the 
product operator $\hat{U}_\sigma(\beta,0)$, restricted to the sub-space of the Hilbert space
spanned by the orbitals $\{\phi_k\}$, i.e.,
\beq
\label{Eq:MDef}
\left[ {M}_\sigma(\beta) \right]^{}_{ab}  = \langle a | \hat{U}_\sigma(\beta,0) | b \rangle .
\eeq
The square of the determinant in Eq.~(\ref{Eq:PsigmaDef}) results from the fact that we are considering
two (distinguishable but otherwise identical) fermion species. Sampling the auxiliary field according 
to $P[\sigma]$, one evaluates the expectation value of observables using
\beq
O_\beta = \frac{1}{\CZ}\int \CD \sigma\, P[\sigma] \, O[\sigma],
\eeq
as a function of the imaginary time $\beta$, where
\beq
O[\sigma] \equiv \frac{\langle \psi^{}_0 |
\,\hat{U}_\sigma(\beta,\beta/2)\,\hat{O}\,\hat{U}_\sigma(\beta/2,0)\,| \psi^{}_0
\rangle}{\langle \psi^{}_0 | \, \hat{U}_\sigma(\beta,0) \,| \psi^{}_0 \rangle}.
\eeq
This is followed by an extrapolation to large $\beta$.

Using this technique, we compute the ground-state energy $E_0$ and Tan's contact
$C$. To determine the ground-state energy, one of the simplest ways to arrive 
at the observable of interest is actually to differentiate $\ln \mathcal Z$, using Eq.~(\ref{Eq:Z}), 
with respect to $\tau$. This is equivalent to using Wick's theorem~\cite{MCReviews}, and it
generates a simple and normalized expression for a stochastic estimator of the energy,
\beq
\label{Eq:EnergyEstimator}
E^{}_\beta = -\frac{\partial \ln \mathcal Z}{\partial \beta} = -\frac{2}{N^{}_\tau}\int \CD \sigma\, P[\sigma] 
\,\tr \left[ {M}^{-1}_\sigma(\beta) \frac{\partial {M}_\sigma(\beta) }{\partial \tau} \right].
\eeq
In the above expression, upon performing the last differentiation, it is easy to identify 
the contribution to $E^{}_\beta$ coming from the kinetic and interaction energies: 
using Eq.~(\ref{Eq:MDef}), the $\tau$ derivative will act on all the factors present in
$\hat U$ and bring down kinetic- or potential-energy terms from the exponent.

Using the estimator for the potential energy, we can calculate Tan's contact $C$, which is defined by
\beq
\label{Eq:Contact}
C \equiv 2 \frac{\partial \langle \hat H \rangle}{\partial a^{}_0} = -g \langle \hat{V}\rangle,
\eeq
where we have used the Feynman-Hellman theorem and the fact that $\hat V$ is just
a contact interaction, which satisfies $g\; {\partial \hat V}/{\partial g} = \hat V$.
The above definition of $C$ holds for the ground state as well as for
excited states. 

In the limit of large $\beta \varepsilon_F$, the difference between the
finite-imaginary-time expectation value $E_\beta$ and the ground-state energy
decays exponentially at a rate determined by the difference between that minimum
energy and the energy of the first excited state $E_1$:
\beq
E_\beta = E_0 + A e^{-\beta(E_1-E_0)} + O\left(e^{-\beta(E_2-E_0)}\right).
\eeq
From this decay, we determine the energy of the first excited state as a function
of the coupling $\gamma$, and proceed to compute the excited state contact $C_1$ via
\beq
\label{Eq:ExcitedContact}
C_1 \equiv 2 \frac{\partial E_1^{}}{\partial a^{}_0} = -g^2 \pderiv{E_1}{g}.
\eeq

To our knowledge, this is the first determination of the contact in an excited state, other than
calculations at finite temperature. In the next section we present our results for these quantities.

%%%%%%%%%%%%%%%%%%%%%%%%%%%%%%%%%%%%%%%%%%%%%%%%%%%%%%%%%%%%
\section{Analysis and Results}

\subsection{Ground- and excited-state energies}

\begin{figure}[t]
\includegraphics[width=1.05\columnwidth]{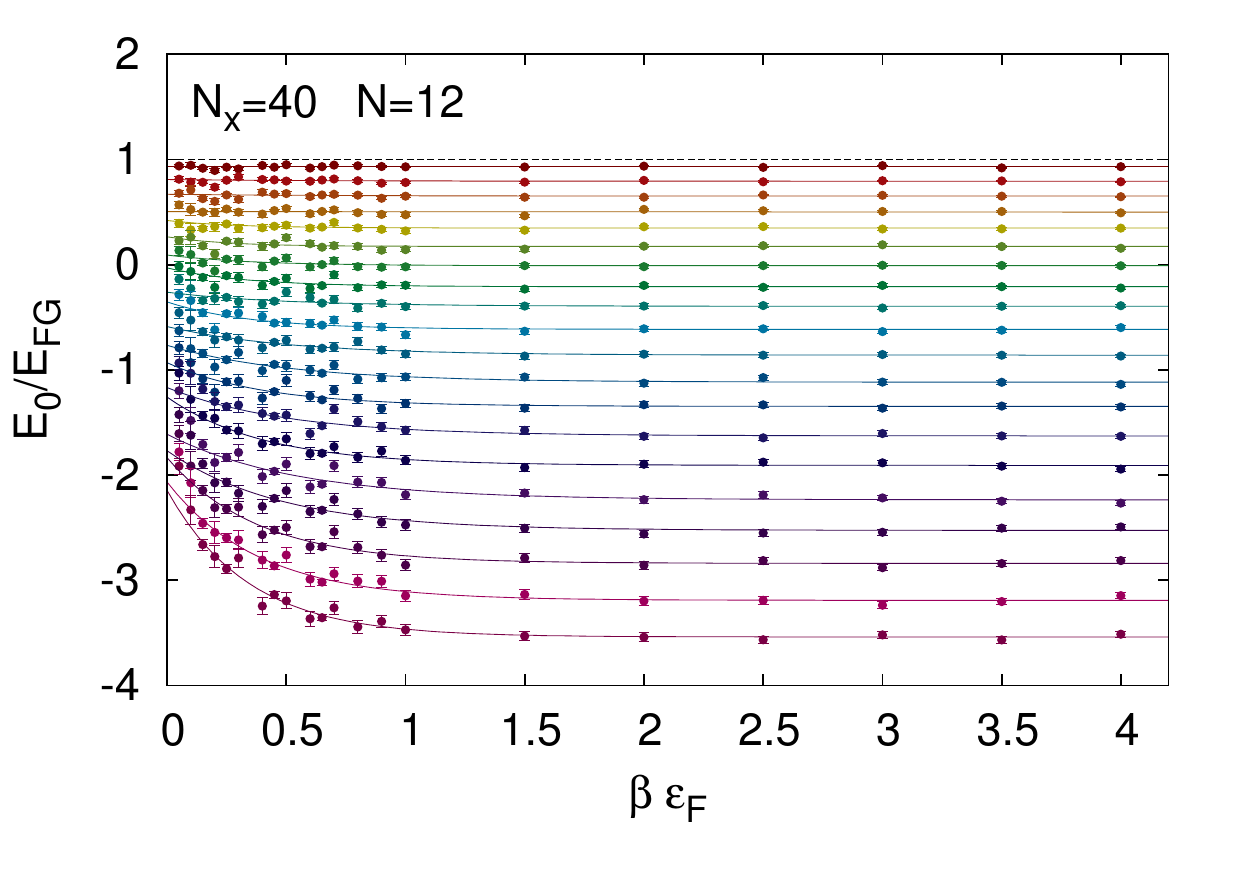}
\includegraphics[width=1.05\columnwidth]{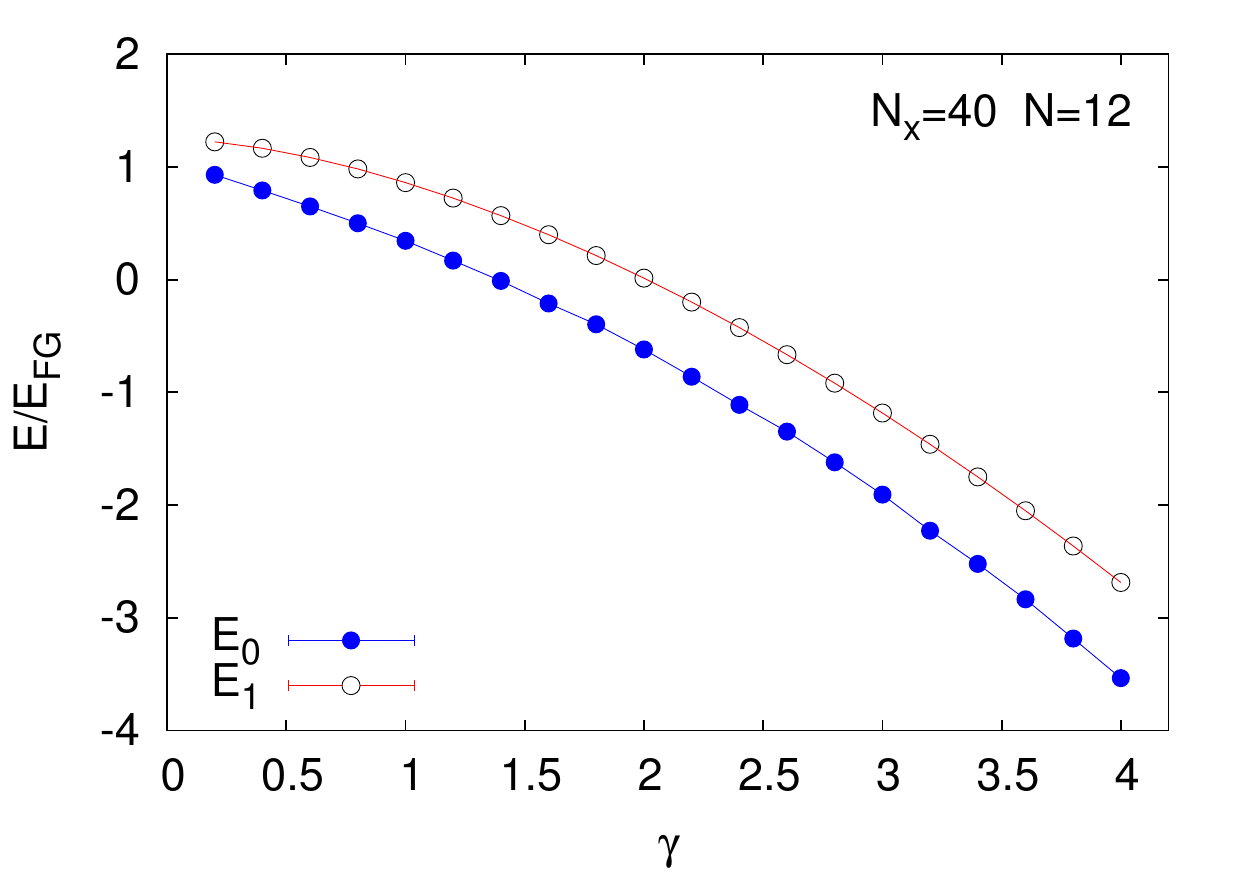}
\caption{\label{Fig:EnergyBetaeF}(color online) Top: Estimate of the energy 
of 12 unpolarized fermions in a $N^{}_x = 40$ lattice with periodic boundary conditions, 
in units of the noninteracting energy $E^{}_\text{FG} = \frac{1}{3} \varepsilon^{}_F N = {\pi^2 N^3}/({24 L^2})$, as a function of imaginary time 
$\beta \varepsilon_F^{}$. From top to bottom, the lines correspond to couplings $\gamma=$ 0, 0.2, \dots, 4.0. The data are shown with statistical error bars. The continuous lines show the
least-squares fits for exponential decay to the ground state.
Bottom: Estimate of the ground- and excited-state energies of 12 unpolarized fermions in 
a periodic 1D box.  The statistical uncertainties are smaller than the size of the symbols.
}
\end{figure}

We performed calculations for $N=4-12$ fermions in lattices of spatial sizes up
to $N_x^{}=40-100$ and couplings $\gamma=$ 0, 0.2, \dots, 4.0. 
We used $\tau=0.025$, which resulted in temporal lattice sizes varying between 60
and up to $10^4$ points.
For each set of parameter values, we took approximately $5000$ decorrelated
samples of the auxiliary field $\sigma$ (which yields statistical uncertainties on the order of 1-2\%)
and evaluated the total-energy estimator of Eq.~(\ref{Eq:EnergyEstimator}). Extrapolating to the
large-$\beta \varepsilon^{}_F$ limit (i.e., large $N_\tau^{}$ limit), we obtained the ground-state values. In
each case, studying the exponential decay of this estimator for increasing imaginary time allowed us to obtain the energy of the first excited state. 
In Fig.~\ref{Fig:EnergyBetaeF} (top panel) we show typical Monte Carlo results for the total energy estimator, with the corresponding fits
used to obtain ground- and excited-state estimates. The error bars represent the statistical uncertainty of the Monte Carlo calculation.

The energy of the ground- and first-excited states, as a function of the coupling $\gamma$, are shown in
Fig.~\ref{Fig:EnergyBetaeF} (bottom panel) for a representative system.
The error bars for the ground-state energy are taken from the exponential-decay fit used to extract the asymptote at large 
imaginary time (top of Fig.~\ref{Fig:EnergyBetaeF}). 
In the same figure we show the first-excited state energy. For the latter, the data were obtained by
studying the exponential decay of Fig.~\ref{Fig:EnergyBetaeF} for all values of $\gamma$ and incorporating, 
using a power-law fit as a function of $\gamma$, the exact value at $\gamma=0$. 
The associated uncertainties represent $95\%$ confidence intervals.

The above lattice results are typical. In order to obtain the physical values in the continuum limit, we extrapolated 
to $N^{}_x \to \infty$. The extrapolated data are shown in Table~\ref{Table:EnergyExtrapolation}, and further details on the extrapolation
procedure are given in Sec.~\ref{Sec:Syst}.

\begin{figure}[t]
\includegraphics[width=1.0\columnwidth]{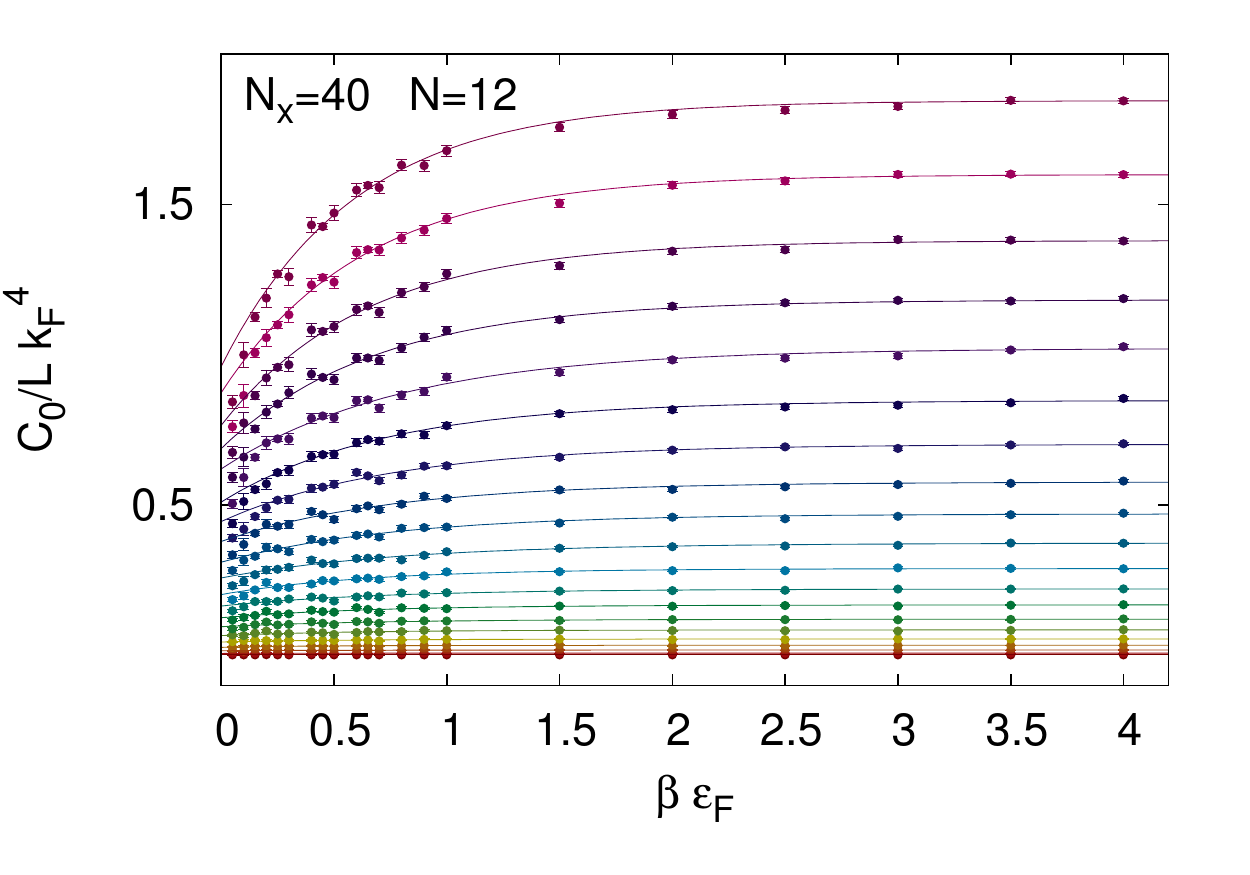}
\includegraphics[width=1.0\columnwidth]{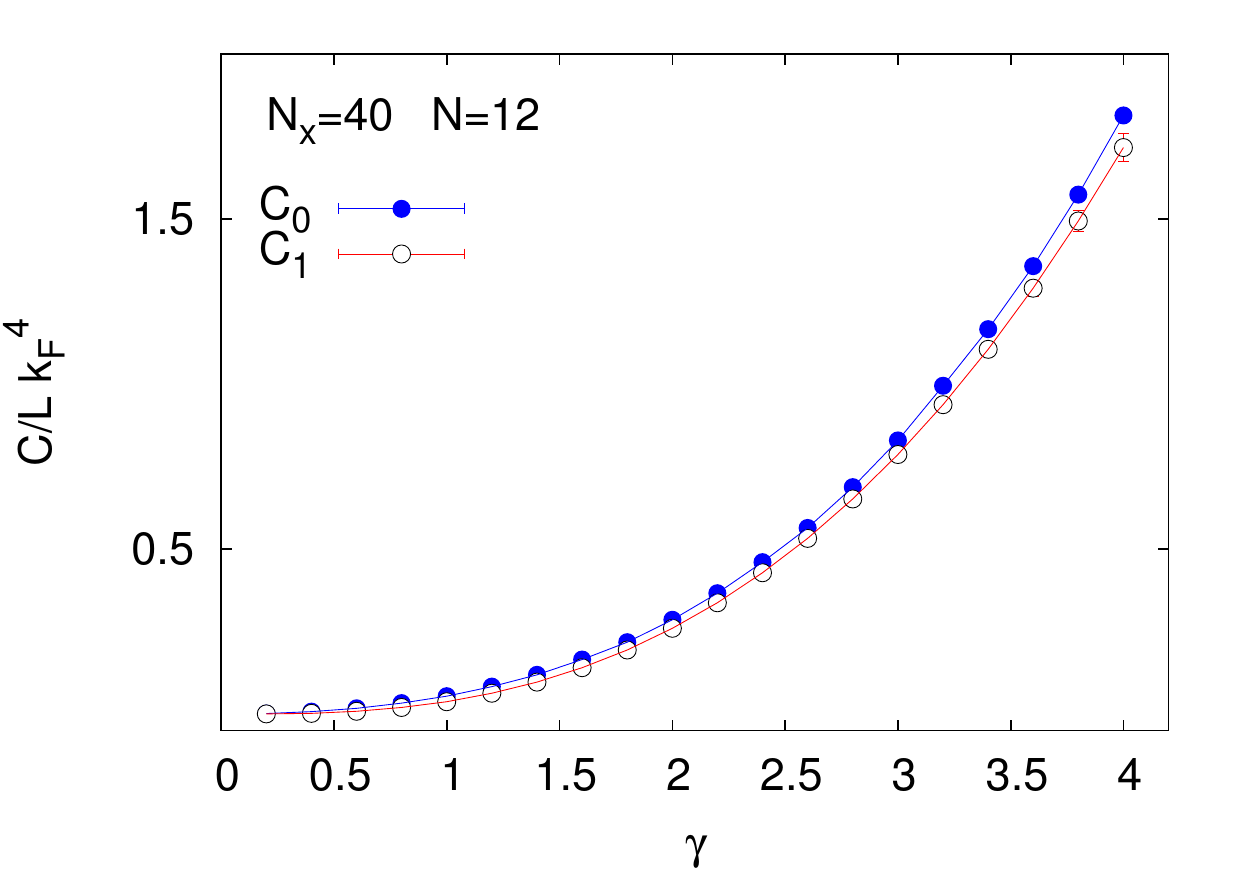}
\caption{\label{Fig:ContactRainbowBeta}(Color online) Top: Estimate of Tan's contact 
in units of $L k_F^4$ for $N = 12$ and $N_x = 40$ for various couplings $\gamma$ showing 
the extrapolation to large $\beta \varepsilon^{}_F$. From bottom to top, the lines correspond to couplings 
$\gamma=$ 0, 0.2, \dots, 4.0. Bottom: Estimates of Tan's
contact for 12 unpolarized fermions in a $N^{}_x = 40$ 
lattice with periodic boundary conditions. The blue line shows the ground-state
result, while the red line shows the excited-state value.
}
\end{figure}
%

%%%%%%%%%%%%%%%%%%%%%%%%%%%%%%%%%%%%%%%%%%%%%%%%%%%%%%%%%%%%
\subsection{Ground- and excited-state contacts}

In order to compute the contact, we proceed as with the energy. We use the 
observable of Eq.~(\ref{Eq:Contact}), which we can access by differentiation of $\ln \mathcal Z$
with respect to the coupling, as explained above. The interaction energy was thus estimated as a 
function of $\beta \varepsilon^{}_F$ and extrapolated to the ground state. The extrapolation procedure is
depicted in Fig.~\ref{Fig:ContactRainbowBeta}.

From the above extrapolation, our results for the contact are shown in the bottom panel of 
Fig.~\ref{Fig:ContactRainbowBeta} for a representative system. The ground-state contact values were taken,
in a similar manner as for the ground-state energy, from the asymptotic behavior of the estimator for the interaction 
energy, and the uncertainties shown were obtained in the same fashion.
The contact associated with the excited state was computed via Eq.~(\ref{Eq:ExcitedContact}). As with the excited-state 
energy, the uncertainties in $C_1^{}$ were obtained by performing a fit to our whole data set as a function of $\gamma$ 
that accounts for the $C_1^{}=0$ value at $\gamma=0$.
Our full results for the contact, as a function of $\gamma$ and $N$, extrapolated to the infinite-volume limit,
are shown in Table~\ref{Table:ContactExtrapolation}.

%%%%%%%%%%%%%
\begin{table*}[t]
\caption{\label{Table:EnergyExtrapolation}
Ground- and excited-state energies ($E_0^{}$ and $E^{}_1$, respectively) as functions of the dimensionless coupling $\gamma$, 
in units of $E^{}_{\text{FG}}$, extrapolated to infinite volume. The uncertainties reported are associated with the extrapolation procedure,
which accounts for the statistical uncertainties in the Monte Carlo calculation at each finite volume.}
\begin{tabularx}{\textwidth}{@{\extracolsep{\fill}} c | c | c c | c c | c c | c c}
\hline\hline
\ \ \ \ $N$\ \ \ \ &\multicolumn{1}{c|}{4} &\multicolumn{2}{c|}{6} &\multicolumn{2}{c|}{8} &\multicolumn{2}{c|}{10} &\multicolumn{2}{c}{12} \\
\hline
$\gamma$   & $E_0^{}$ & $E_0^{}$ & $E_1^{}$ & $E_0^{}$ & $E_1^{}$& $E_0^{}$ & $E_1^{}$& $E_0^{}$ & $E_1^{}$\\
\hline
0.00  &   	1.5  		&   	0.888\dots&	1.555\dots	&  1.125  		&	1.59375		&   	0.96  		&	1.2	 	&   	1.0555\dots  &	1.25 \\
0.20  &   	1.34(1)   	&   	0.771(6) 	&	 1.542(3)	&  0.99(2) 		&	1.58(3)  	& 	0.83(3)   		&	1.19(1)    	&   	0.923(4)  	&	1.23(1) \\
0.40  & 	1.22(1)  	&   	0.63(1)	&	1.50(1) 	&  0.852(6)	&	1.54(7)   	&	0.69(7)		& 	1.14(3)    	&	0.794(2)	&	1.18(2) \\
0.60  &   	1.07(1)  	& 	0.51(2)	& 	1.43(2)	&  0.711(1)	&	1.5(1)   	&	0.55(1)		&	1.08(6)    	&	0.649(5)	&	1.10(4) \\
0.80  &   	0.94(4)  	& 	0.33(2)	&	1.33(3) 	&  0.542(5)	&	1.4(1)   	&	0.40(2)		&	0.99(8)   	&	0.49(1)	&	1.00(6) \\
1.00  &   	0.67(3)  	& 	0.17(2)	&	1.21(4) 	&  0.37(2)		&	1.2(2)   	&	0.23(2)		&	0.9(1)   	&	0.31(1)	&	0.86(7) \\
1.20  &   	0.50(2)  	& 	-0.04(2)	&	1.07(5) 	&  0.18(1)		&	1.1(2)   	&	0.06(1)		&	0.7(1)   	&	0.11(3)	&	0.71(8) \\
1.40  &   	0.28(3)  	& 	-0.19(1)	&	0.91(6) 	&  -0.01(2)	&	0.9(2)  	&	-0.17(1)		&	0.6(2)   	&	-0.06(1)	&	1.0(9) \\
1.60  &   	0.10(2)  	& 	-0.42(1)	&	0.73(7) 	&  -0.27(1)	&	0.7(2)  	&	-0.37(1)		&	0.4(2)    	&	-0.30(2)	&	0.3(1)  \\
1.80  &   	-0.2(1)  	& 	-0.65(3)	&	0.52(8)	&  -0.50(4)	&	0.5(2)  	&	-0.61(3)		&	0.2(2)   	&	-0.55(2)	&	0.1(1) \\
2.00  &     -0.51(5)  	& 	-0.88(3)	&	0.30(9) 	&  -0.75(2)	& 	0.3(2) 	&	-0.82(2)		&	 -0.04(20)	&	-0.79(2)	&	 -0.1(1)\\
2.20  &   	-0.85(4)  	&	-1.11(2)	&	0.06(9)	&  -1.05(3)	&	0.09(20)	&	-1.07(1)		&	 -0.3(2) 	&	-1.03(4)	&	 -0.4(1)\\
2.40  &  	-0.94(7)  	&	-1.41(3)	&	-0.19(9)	&  -1.32(1)	&	-0.2(2)	&	-1.37(2)		&	 -0.5(2) 	&	-1.34(2)	&	 -0.7(1)\\
2.60  &   	-1.43(3) 	&	-1.76(4)	&	-0.5(1)	&  -1.62(1)	&	 -0.4(2) 	&	-1.68(2)		&	 -0.8(2) 	&	-1.65(1)	&	 -1.0(1)\\
2.80  &   	-1.79(6)  	&	-2.08(4)	&	-0.7(1)	&  -1.98(3)	&	 -0.7(1)	&	-2.01(5)		&	 -1.1(2)   	&	-1.98(1)	&	 -1.3(1) \\
3.00  &   	-2.2(1)  	&	-2.37(7)	&	-1.0(1)	&  -2.35(1)	&	 -0.9(1)	&	-2.39(2)		&	 -1.5(2)  	&	-2.33(2)	&	 -1.7(1) \\
3.20  &   	-2.3(1)  	&	-2.83(2)	&	-1.4(1)	&  -2.68(1)	&	 -1.2(1) 	&	-2.73(2)		&	 -1.8(2)   	&	-2.69(3)	&	 -2.0(1) \\
3.40  &   	-2.9(1) 	&	-3.16(3)	&	-1.7(1)	&  -3.07(1)	&	 -1.49(1) 	&	-3.09(4)		&	 -2.2(2)  	&	-3.16(4)	&	 -2.4(1) \\
3.60  &   	-3.4(2)  	&	-3.55(2)	&	-2.0(2)	&  -3.54(3)	&	 -1.8(1) 	&	-3.56(1)		&	 -2.6(1) 	&	-3.53(4)	&	 -2.8(1) \\
3.80  &   	-3.8(1)  	&	-4.05(2)	&	-2.3(2)	&  -3.95(7)	&	 -2.2(2)  	&	-4.00(4)		&	 -3.03(10)  	&	-3.96(4)	& -3.2(1)\\
4.00  &   	-4.2(1)  	&	-4.51(7)	&	-2.6(3)	&  -4.38(5)	&	 -2.5(3) 	&	-4.49(4)		&	 -3.45(6)  	&	-4.39(7)	&	 -3.6(2) \\
\hline\hline
\end{tabularx}
\end{table*}
%%%%%%%%%%%%%
%%%%%%%%%%%%%
\begin{table*}[t]
\caption{\label{Table:ContactExtrapolation}
Ground- and excited-state contacts ($C_0^{}$ and $C^{}_1$, respectively) as functions of the dimensionless coupling $\gamma$, 
in units of $L k_F^4$, extrapolated to infinite volume. The uncertainties reported are associated with the extrapolation procedure,
which accounts for the statistical uncertainties in the Monte Carlo calculation at each finite volume.}
\begin{tabularx}{\textwidth}{@{\extracolsep{\fill}} c | c | c c | c c | c c | c c}
\hline\hline
\ \ \ \ $N$\ \ \ \ &\multicolumn{1}{c|}{4} &\multicolumn{2}{c|}{6} &\multicolumn{2}{c|}{8} &\multicolumn{2}{c|}{10} &\multicolumn{2}{c}{12} \\
\hline
$\gamma$   & $C_0^{}$ & $C_0^{}$ & $C_1^{}$ & $C_0^{}$ & $C_1^{}$& $C_0^{}$ & $C_1^{}$& $C_0^{}$ & $C_1^{}$\\
\hline
0.00  &   	0  			&   	0			&	0 			&  0  			&	0 			&   	0 	 		&	0 		&   	 0  			&	0 		\\
0.20  &   	0.00221(7)  	&   	0.00167(7)	&	0.00023(4)	&  0.00186(3)	&	0.0006(4) 		& 	0.00179(4)	&	0.0002(1) 	&   	 0.00185(5)	&	0.0003(1) 	\\
0.40  & 	0.0081(3) 		&   	0.0073(4)		&	0.0018(3)	 	&  0.0079(2)	&	0.003(2)  		&	0.0076(2)		& 	0.0014(7) 	&	 0.00750(4)	&	0.0020(8)  \\
0.60  &   	0.0189(3) 		& 	0.0169(8)		& 	0.007(1) 		&  0.0185(2)	&	0.007(5)  		&	0.0179(5)		&	0.006(3) 	&	 0.0182(2)		&	0.008(2)  	\\
0.80  &   	0.034(2) 	 	& 	0.034(1)		&	0.018(2)  		&  0.0357(4)	&	0.017(9)  		&	0.034(1)		&	0.017(6) 	&	 0.0344(9)		&	0.020(3)  	\\
1.00  &   	0.063(2)	 	& 	0.056(2)		&	0.036(4)	 	&  0.060(1)	&	0.03(1)  		&	0.058(1)		&	0.03(1) 	&	 0.0594(8)		&	0.040(5)  	\\
1.20  &   	0.094(3)	 	& 	0.089(2)		&	0.063(5)	 	&  0.092(1)	&	0.06(2) 	 	&	0.088(1)		&	0.06(1) 	&	 0.091(2)		&	0.069(7)  	\\
1.40  &   	0.138(3)  		& 	0.127(1)		&	0.100(7) 	 	&  0.132(2)	&	0.10(2)  		&	0.131(1)		&	0.10(2) 	& 	 0.132(1)		&	0.109(9)  	\\
1.60  &   	0.188(4)	  	& 	0.180(1)		&	0.148(8) 	 	&  0.195(1)	&	0.14(2)  		&	0.184(1)		&	0.15(2) 	& 	 0.186(2)		&	0.16(1)  	\\
1.80  &   	0.26(2)	 	& 	0.243(3)		&	0.21(1) 	 	&  0.260(6)	&	0.20(1)  		&	0.249(3)		&	0.21(2) 	&	 0.254(3)		&	0.23(1)  	\\
2.00  &      0.35(1)	  	& 	 0.317(3)		&	 0.28(1) 		&   0.342(3)	& 	0.264(4) 		&	 0.326(4) 		&	  0.29(2) 	&	 0.335(2)		&	0.31(1) 	\\
2.20  &   	 0.45(1)	  	&	 0.411(8)		&	 0.37(2) 		&   0.448(6)	&	0.344(8) 		&	 0.423(1)		&	  0.38(2) 	&	 0.434(5)		&	0.41(1) 	\\
2.40  &  	 0.53(1)	  	&	 0.530(9)		&	  0.47(2) 	  	&   0.566(1)	&	0.44(2)  		&	 0.543(2)		&	  0.50(2)  	&	 0.551(3)		&	0.53(2)  	\\
2.60  &   	 0.72(3)	 	&	 0.67(1)		&	  0.60(3) 		&   0.705(2)	&	0.54(5) 		&	 0.684(5)		&	  0.63(1) 	&	 0.693(1)		&	0.67(2) 	\\
2.80  &   	 0.90(1)	  	&	 0.83(1)		&	  0.73(5) 		&   0.879(7)	&	0.66(8) 		&	 0.852(8)		&	  0.79(1) 	&	 0.857(2)		&	0.83(3) 	\\
3.00  &   	 1.07(4)	 	&	 1.00(2)		&	  0.88(7) 		&   1.072(1)	&	0.8(1) 		&	 1.04(1)		&	  0.97(2) 	&	 1.046(3)		&	1.01(4) 	\\
3.20  &   	 1.19(4)	 	&	 1.26(2)		&	  1.05(9) 		&   1.287(3)	&	0.9(2) 		&	 1.26(1)		&	  1.17(5) 	&	 1.26(1)		&	1.21(6) 	\\
3.40  &   	 1.52(4) 		&	 1.48(2)		&	  1.2(1) 	 	&   1.532(4)	&	1.1(2)   		&	 1.51(1)		&	  1.40(8)  	&	 1.51(2)		&	1.44(7)  	\\
3.60  &   	 1.80(5) 		&	 1.78(1)		&	  1.4(1) 		&   1.83(1)		&	1.4(3)  		&	 1.80(1)		&	  1.7(1) 	&	 1.79(2)		&	1.7(1)  	\\
3.80  &   	 2.06(7)	 	&	 2.10(3)		&	  1.6(2) 		&   2.14(2)		&	1.6(4)  		&	 2.09(2)		&	  1.9(2)	&	 2.09(1)		&	2.0(1)  	\\
4.00  &   	 2.40(5) 	  	&	 2.47(2)		&	  1.9(2) 	  	&   2.45(1)		&	1.9(5)   		&	 2.47(1)		&	  2.2(2)  	&	 2.44(4)		&	2.3(2)  	\\
\hline\hline
\end{tabularx}
\end{table*}
%%%%%%%%%%%%%

%%%%%%%%%%%%%%%%%%%%%%%%%%%%%%%%%%%%%%%%%%%%%%%%%%%%%%%%%%
%%%%%%%%%%%%%%%%%%%%%%%%%%%%%%%%%%%%%%%%%%%%%%%%%%%%%%%%%%%%
\subsection{Approach to the thermodynamic limit}

\begin{figure}[t]
\includegraphics[width=0.923\columnwidth]{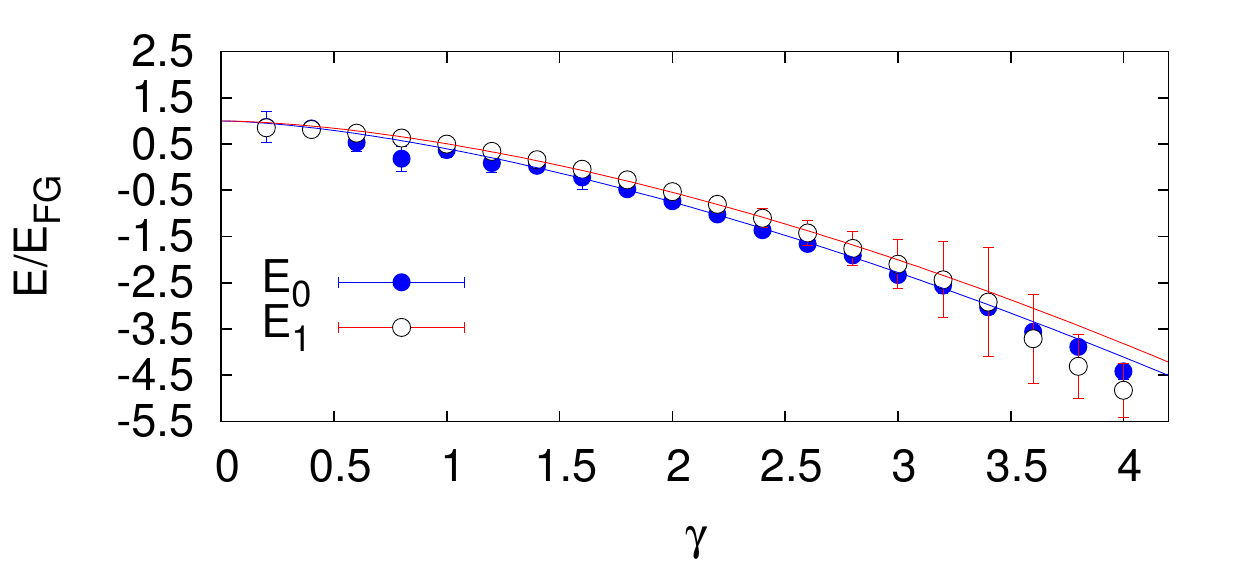}
\includegraphics[width=0.923\columnwidth]{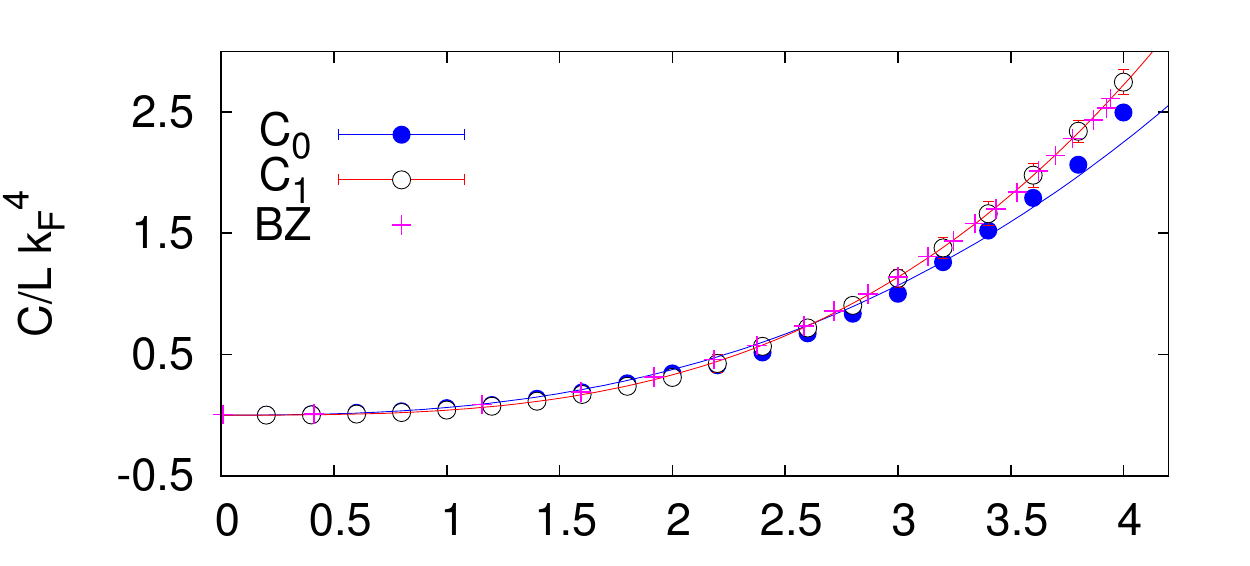}
\caption{\label{Fig:ThermoLimit}(Color online) Ground- and excited-state energy (top) and contact (bottom) as functions 
of the dimensionless coupling $\gamma$, extrapolated to the thermodynamic limit. 
The crosses (indicated as BZ in the legend) show the thermodynamic-limit data from Ref.~\cite{BarthZwerger}, and the solid lines are fits (see text).
}
\end{figure}

As can be appreciated in Tables~\ref{Table:EnergyExtrapolation} and~\ref{Table:ContactExtrapolation}, the variation in $E/E^{}_\text{FG}$ 
and $C/(L k_F^4)$ as a function of particle number
is relatively small. This behavior is indicative of a rather fast approach to the thermodynamic limit. To quantify this feature more precisely, we extrapolate our results to the large-$N$ limit. Note that we are approaching that limit by first taking the large-volume 
limit, i.e., the thermodynamic limit is reached along the line of vanishing density. This is the preferred path for lattice calculations 
in order to avoid finite-range effects. Our results are shown in Fig.~\ref{Fig:ThermoLimit}, where the fits
are of the form
\beq
f(\gamma) = f^{}_0 + A\gamma^B,
\eeq
and where $f^{}_0 = 1$ for the energy fits and 0 for the contact fits.
We find $A= -0.60(2)$, $B=1.54(4)$ for the ground-state energy; $A=-0.50(2)$, $B=1.64(4)$ for the excited-state energy; 
$A=0.063(3)$, $B=2.58(4)$ for the ground-state contact; and $A=0.041(1)$, $B=3.04(1)$ for the excited-state contact.
In the thermodynamic limit, the system is known to be gapless, such that we expect our extrapolations of $E^{}_0$ and $E^{}_1$
to coincide; they do so only up to $\gamma \simeq 3.0$ within the extrapolation uncertainties.
In the same figure, our answers are compared with those obtained directly in the thermodynamic limit in Ref.~\cite{BarthZwerger},
which uses the Bethe ansatz. The agreement is not perfect but quite satisfactory, especially for $\gamma \leq 3.0$.

%%%%%%%%%%%%%%
%

%%%%%%%%%%%%%%%%%%%%%%%%%%%%%%%%%%%%%%%%%%%%%%%%%%%%%%%%
\section{\label{Sec:Syst}Systematic effects}

%%%%%%%%%%%%%%
\begin{figure}[t]
\includegraphics[width=0.923\columnwidth]{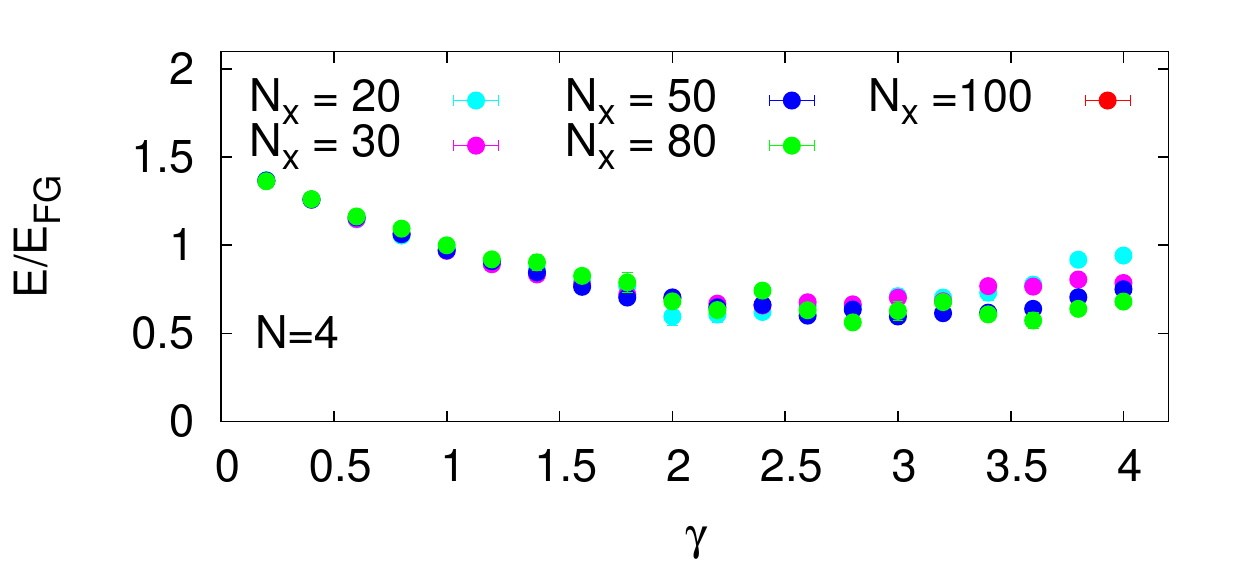}
\includegraphics[width=0.923\columnwidth]{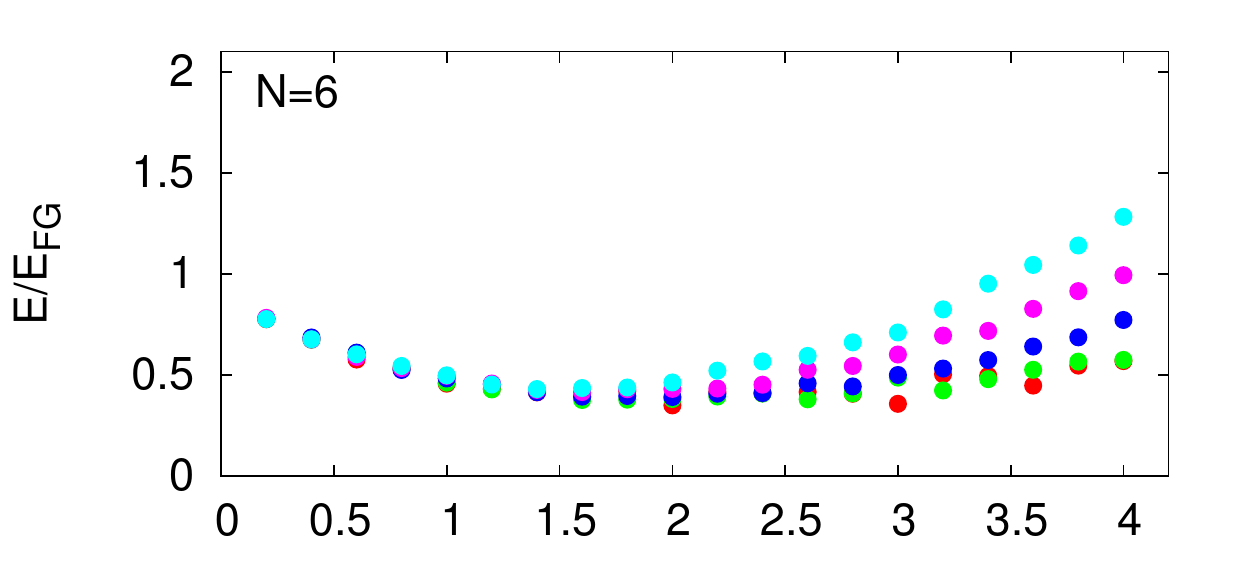}
\includegraphics[width=0.923\columnwidth]{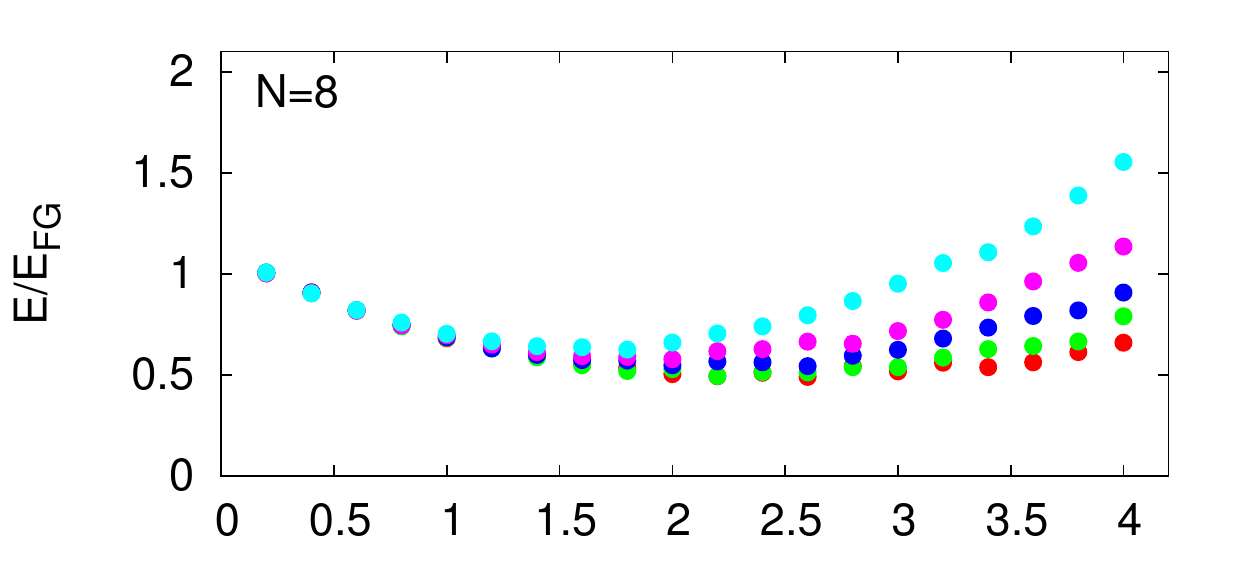}
\includegraphics[width=0.923\columnwidth]{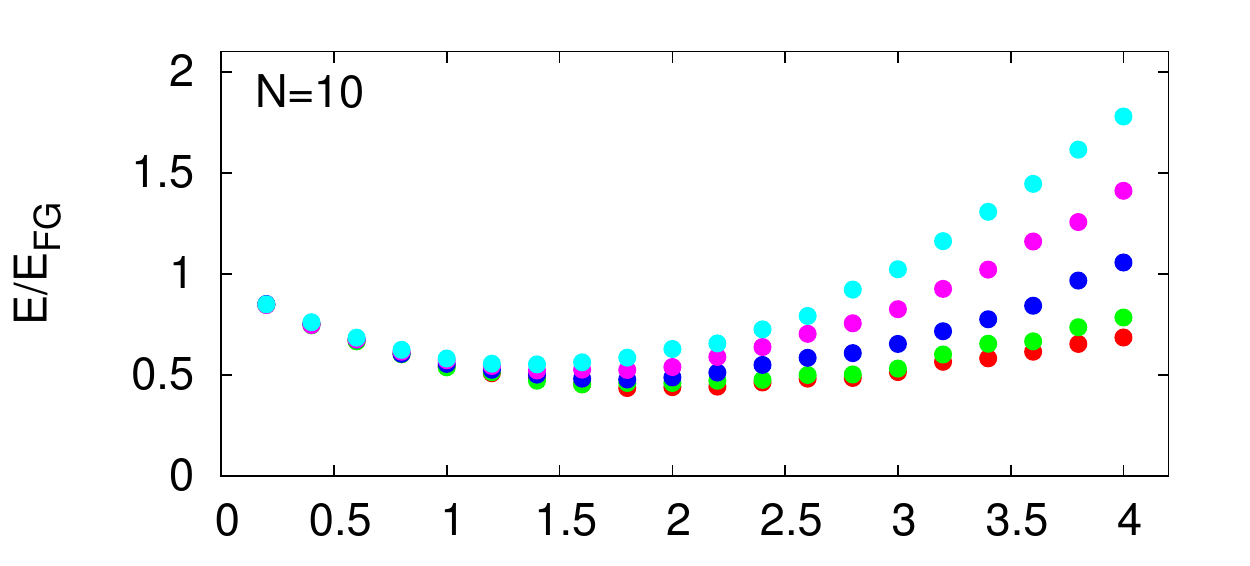}
\includegraphics[width=0.923\columnwidth]{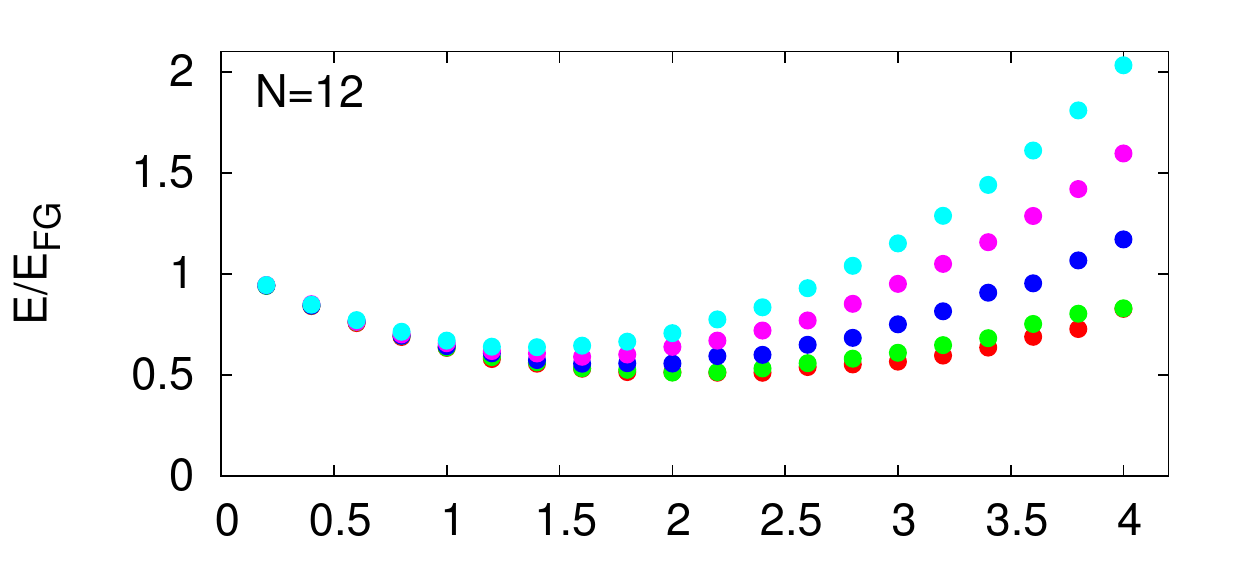}
\caption{\label{Fig:EnergyVolGammaAllN}(Color online) Ground-state energy of $N=$ 4, 6, \dots, 12 particles (top to bottom panels, respectively) 
in a periodic 1D box with volumes $N^{}_{x} =$ 20, 30, 50, 80, and 100. The binding energy of $N/2$ pairs, i.e. $E^{}_\text{B} = N \varepsilon^{}_B/2$ 
was added, where $E^{}_\text{B}/E^{}_\text{FG} = 3\gamma^2/\pi^2$ and $\varepsilon^{}_B = 1/a_0^{2} = g^2/4$.}
\end{figure}

In this section we elaborate on the systematic effects due to finite lattice size.
At fixed particle number, increasing the lattice volume reduces the density. Thus, lattice-spacing effects
are systematically reduced as the interparticle spacing grows well beyond the lattice spacing. 
The effect of increasing the lattice size is clear in Fig.~\ref{Fig:EnergyVolGammaAllN}, where we show the
energy shifted by the binding energy, which yields the energy of $N/2$ (effective) bosonic dimers:
\beq
\frac{E^{}_{\text{eff},N/2}}{E^{}_{\text{FG}}} = \frac{E^{}_{N}}{E_{\text{FG}}}+\frac{3\gamma^2}{\pi^2}.
\eeq
As evident in that figure, the change due to increased lattice size is monotonic and essentially converged at $N^{}_x$=100 to within 
the statistical uncertainty of our calculations. A slower rate of convergence is nevertheless observed as $N$ is increased.
The fact that a single universal curve is achieved for each value of $N$
indicates that the continuum limit is reached and is consistent with $\gamma=g/n$ being the correct physical
coupling. In order to extrapolate to the continuum limit, we found it sufficient to perform a linear fit to our data as a function
of $1/N_x^{}$ at constant $N$ and $g$. 

The top panel of Fig.~\ref{Fig:EnergyVolGammaAllN} ($N$=4) also gives insight into the nature of the effective dimer-dimer interaction,
which has been studied before with the aid of the Bethe ansatz and
other methods (see e.g. Refs.~\cite{AstrakharchikBlume,BatchelorEtAl}). It is well known that, in the strong-attraction limit,
the fermion pairs become effectively hard-core bosons~\cite{GaudinYang}. As we see in Fig.~\ref{Fig:EnergyVolGammaAllN}, 
the effective boson interaction must be at least partially attractive in character (in that increasing the coupling lowers the total energy) up to $\gamma\simeq 2.2$, and 
saturates or becomes mild and fully repulsive beyond that point.

\begin{figure}[t]
\includegraphics[width=1.0\columnwidth]{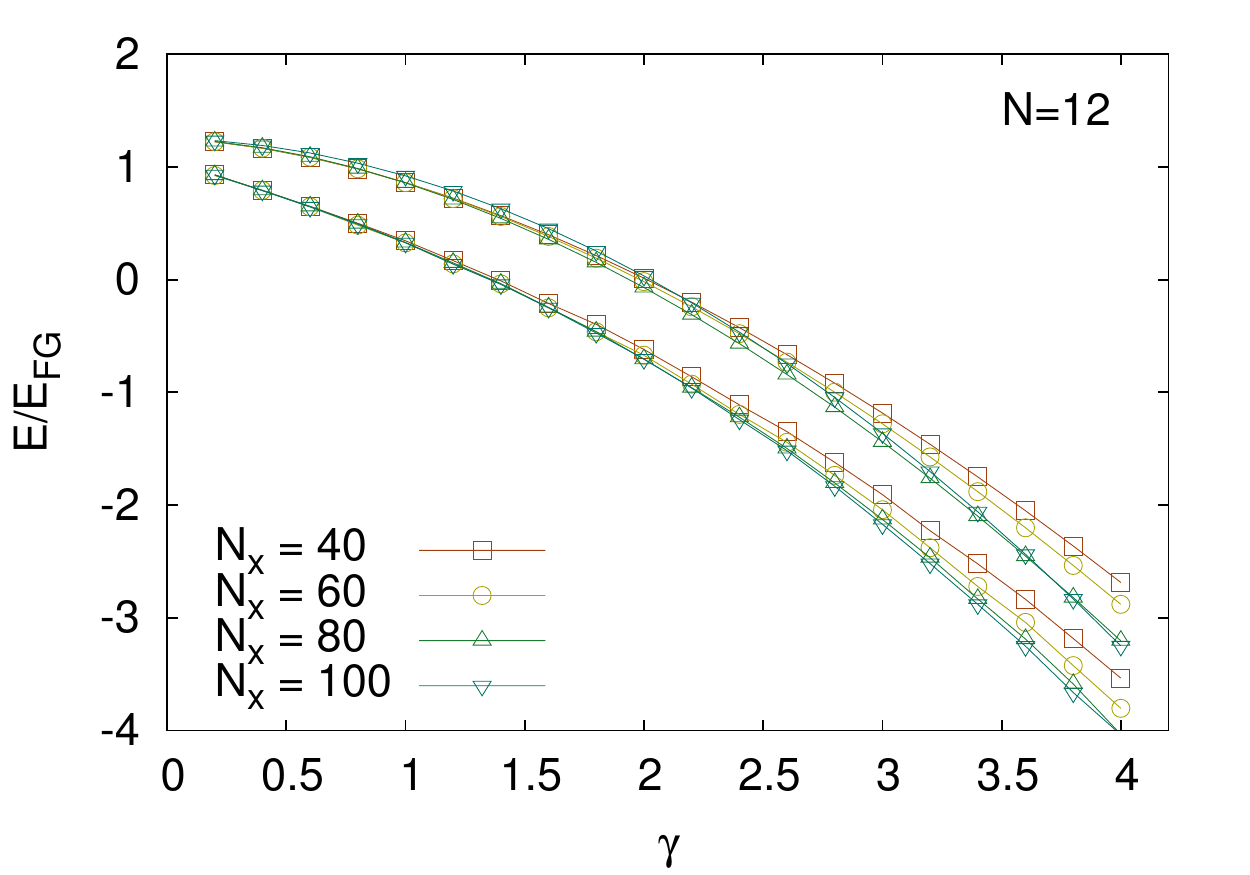}
\caption{\label{Fig:EnergyVolGammaN12}(Color online) Estimates of the ground- and excited-state energies of a 12-particle system in 
a periodic 1D box with volumes $N^{}_{x} = 40$, 60, 80, and 100.}
\end{figure}

While the behavior of the ground-state energy with the lattice size is rather benign, the excited-state energies are affected in
a more pronounced way, as shown in Fig.~\ref{Fig:EnergyVolGammaN12}. As can be appreciated in that figure, 
the lattice effects are larger, which makes their extrapolation to the infinite-volume limit more challenging.

%%%%%%%%%%%%%%

%%%%%%%%%%%%%%%%%%%%%%%%%%%%%%%%%%%%%%%%%%%%%%%%%%%%%%%%
\section{Summary and conclusions} 

We have presented lattice Monte Carlo results for the ground- and excited-state energies as well as
the ground- and excited-state contacts for few-fermion systems in
a periodic one-dimensional box. 
Our calculations were performed on lattices of size $N_x^{}=$ 20, \dots, 100 and are exact up to statistical and 
systematic uncertainties, both of which we have addressed explicitly. 
Our results cover unpolarized systems with particle numbers in the range
$N=$ 4 -- 12, and attractive coupling strengths 
$\gamma = 0$, 0.2, \dots, 4.0. Although these systems can be studied directly with the Bethe
ansatz in the continuum and thermodynamic limits, we provide here results for finite systems, 
spanning the few- to many-body regimes and showing explicitly 
the approach to the thermodynamic limit for the energy and the contact. Our results
indicate that that limit is approached surprisingly quickly.

When analyzing the numerical data for excited-state quantities, we found it necessary to apply a simple yet powerful technique to
determine the uncertainties in a realistic fashion. This approach consisted of performing power-law fits that took into account the exact
results in the noninteracting limit. Such a constraint, together with the assumption that the underlying curve
was smooth, allowed us to provide reasonable and well-defined estimates for the uncertainties for $E_1^{}$ and $C_1^{}$, except for the
$N=4$ case, which proved numerically more challenging. We defer further study of that case to future work.

%From our power-law fits we extracted the critical coupling $\gamma_c^{}$ for many-body bound-state formation for each $N$ 
%considered, and compared them with a strong-coupling approximation. We found good agreement only at large $N$.

Where possible, we compared our answers with previous results in the thermodynamic limit (e.g., from Ref.~\cite{BarthZwerger})
and found very good agreement.  Further, our results serve as a benchmark for higher-dimensional studies that make use of the 
lattice Monte Carlo method to study few-fermion systems at zero temperature.

Our calculations provide a picture of the effective interaction between bosonic pairs in the $N\geq4$ system.
The nature of that interaction depends on the coupling strength; it lowers the energy relative to the total 
binding energy up to $\gamma\simeq 2.2$, and saturates or becomes weakly repulsive beyond that.

Finally, it should be stressed that the limitations we set in our study (namely a maximum particle number $N=12$ and lattice size 
$N_x^{}=100$) do not reflect a breakdown of the method of any kind. Higher particle numbers and stronger couplings
do require larger lattices in order to achieve the continuum limit, but this can also be reached confidently by extrapolation
from smaller systems and lattices, as we show here. However, a better route may be the use of improved actions and operators 
(see, e.g., Ref.~\cite{ImprovedActionsDrut}) to reduce lattice spacing effects at fixed volume.

%%%%%%%%%%%%%%%%%%%%
\acknowledgements
This material is based upon work supported by the 
National Science Foundation Nuclear Theory Program under Grant No. PHY{1306520},
and the
National Science Foundation Graduate Research Fellowship Program under Grant No. DGE{1144081}.

%%%%%%%%%%%%%%%%%%%%%%%%%%%%%%%%%%%%%%%%%%%%%%%%%%%%%%

%%%%%%%%%%%%%%%%%%%%%%%%%%%%%%%%%%%%%%%%%%%%%%%%%%%%%%

\end{document}